\documentclass[aps,prl,twocolumn,showpacs,preprintnumbers,amsmath,amssymb, superscriptaddress]{revtex4}
\usepackage{graphicx}
\usepackage{dcolumn}
\usepackage{bm}
\usepackage{color, float}
 \usepackage{comment}

\usepackage{subfigure}

\usepackage{multirow}

\usepackage{amssymb}
\usepackage{amsmath}
\def\ah{\hat{a}_{\rm H}}
\def\ahd{\hat{a}^{\dagger}_{\rm H}}
\def\av{\hat{a}_{\rm V}}
\def\avd{\hat{a}^{\dagger}_{\rm V}}

\begin{document}

\preprint{xxx}

\title
{
Quantum-enhanced phase estimation using optical spin squeezing
}

\author{Takafumi Ono
}
\email{Takafumi.Ono@Bristol.ac.uk}
\author{Javier Sabines Chesterking
}
\author{Hugo Cable
}
\author{Jeremy L. O'Brien
}
\author{Jonathan C. F. Matthews
}

\affiliation{%
Quantum Engineering Technology Labs, H. H. Wills Physics Laboratory and Department of Electrical $\&$ Electronic Engineering, University of Bristol, BS8 1FD, UK.
}%

\date{\today}
%

\begin{abstract}
\noindent
Quantum metrology enables estimation of optical phase shifts with precision beyond the shot-noise limit.
One way to exceed this limit is to use squeezed states,
where the quantum noise of one observable is reduced at the expense of increased quantum noise
for its complementary partner. 
Because shot-noise limits the phase sensitivity of all classical states, 
reduced noise in the average value for the observable being measured allows for improved phase sensitivity.
However, additional phase sensitivity can be achieved using phase estimation strategies that account for the full distribution of measurement outcomes.
Here we experimentally investigate the phase sensitivity of a five-particle optical spin-squeezed state generated by photon subtraction from a parametric downconversion photon source. The Fisher information for all photon-number outcomes shows it is possible to obtain a quantum advantage of 1.58 compared to the shot-noise limit, even though due to experimental imperfection, the average noise for the relevant spin-observable does not achieve sub-shot-noise precision. Our demonstration implies improved performance of spin squeezing for applications to quantum metrology.
\end{abstract}

\pacs{07.05.Kf, 03.65.Wj, 03.67.-a, 42.50.Dv, 42.50.Ex}
\maketitle
%
\section{Introduction}
Quantum metrology uses non-classical states to enable measurement of physical parameters with precision beyond the fundamental shot-noise limit \cite{Giovannetti2011}. This is subject to intense research effort for measurements at the single-photon level \cite{He2013}, with higher intensity quantum optics \cite{Vahlbruch2016} and with matter \cite{Zoest2010}. In all these cases the central motivation is to understand how to extract more information per-unit of resource (such as probe power and interaction time) and this will naturally lead to applications in precision measurement \cite{Taylor2013, Aasi2013, Bloom2014, Hosten2016}. An approach that dates back to the beginning of quantum optics \cite{Caves1981} is to improve phase sensitivity using squeezed states \cite{Vahlbruch2016}. In discrete quantum optics, one approach has been to use path-entangled states, such as NOON states \cite{Dowling2008}, as a means to achieve supersensitivity since they exhibit interference patterns with increased frequency compared to classical light. So far, experiments have reached photon numbers of up to six \cite{Nag07, Afek2010, Xiang2013} photons, and recent works aim to address weaknesses in these schemes due to loss \cite{Kacprowicz2010} and state generation using non-deterministic processes \cite{Jonathan2013}. Using probes multiple times can also enable a precision advantage, which varies according to the chosen notion of resource \cite{Higgins2007, Birchall2016}.

Spin squeezing has proven to be a useful approach thanks to developments in experiments manipulating atomic ensembles \cite{Luis2006,Ono2008,Shalm2009,Beduini2013,Rozema2014,Beduini2015}. In these experiments ensemble measurements are typically used, which correspond to collective observables for all particles in the ensemble. However, experiments that utilise detections at the single-particle level, can in principle achieve sensitivity beyond that achievable using ensemble measurements \cite{Luca2009}. The total statistical information that can be extracted from a measurement of an unknown phase shift is captured by the Fisher information \cite{Helstrom1976, Fisher1925}, which is evaluated for all measurement outcomes. Because it is well known that squeezing can improve the phase sensitivity in many set-ups, it is important to quantify the sensitivity improvement with squeezing, and how close this sensitivity is to the maximum phase sensitivity as quantified using Fisher information.

In this paper, we focus on measurements using spin squeezing \cite{Kitagawa1993, Ma2011}, which has been shown to enable supersensitive precision in several experiments using ultracold atoms \cite{Appel2009,Gross2010,Lucke2011,Hamley2012}. We report on an experimental investigation of the phase sensitivity achievable using an optical spin-squeezed state, which was originally considered by Yurke et. al \cite{Yurke1986, Lee02}. Our setup generates five-photon Yurke states using one-photon subtraction from light emitted by a parametric downconversion source \cite{Hofmann2006}, and we use spatially multiplexed pseudo-number-resolving detection to reconstruct photon-number statistics at the output \cite{Jonathan2013}. Our analysis demonstrates supersensitive phase measurement from the observed optical Yurke state, using all five-photon coincidence outcomes. We investigate the role of spin squeezing in achieving this quantum enhancement.

\section{Theory}
Consider first $N$ uncorrelated single photons, 
where each photon is in a superposition of horizontal (H) and vertical (V) polarizations, 
$(|1, 0 \rangle_{\rm HV} + |0, 1 \rangle_{\rm HV} )/\sqrt{2}$.
When we measure this state in HV-polarization basis, 
the probability that $n$ photons are detected with H polarization and $N-n$ photons 
 with V polarization is given by the Binomial distribution $P_{n} = \binom {N} {n} (1/2)^N$.
The noise obtained from this distribution is given by $\sqrt{N}$, which is called shot noise
for phase estimation.

The state which we consider here, sometimes referred to as the Yurke state \cite{Yurke1986},  is a superposition of the two states of 
$(N+1)/2$ photons are in one optical mode (e. g. horizontal polarization) 
and $(N+1)/2-1$ photons are in an orthogonal mode (e. g. vertical polarization) of the form 
$(|(N-1)/2,(N+1)/2 \rangle+|(N+1)/2,(N-1)/2 \rangle)/\sqrt{2}$, where $N \geq 3$ is restricted to odd values. 
If we adopt polarization encoding and measure in the HV basis, the outcomes take two values with photon-number difference $\pm 1$. The noise of Yurke state is therefore $1$ which is smaller than that using 
$N$ uncorrelated photons with a noise of $\sqrt{N}$.

More generally, the photon statistics of any two-mode $N$-photon system can be described by
the Stokes parameters describing the photon-number differences
 between H and V polarization, 
 diagonal (D)  and anti-diagonal (A) polarization, and right-circular (R) and left-circular (L) polarization,
\begin{eqnarray}
\hat{S}_1 &=& \hat{n}_{\rm H} - \hat{n}_{\rm V} = \ahd \ah - \avd \av \nonumber\\
\hat{S}_2 &=& \hat{n}_{\rm D} - \hat{n}_{\rm A} = \ahd \av + \avd \ah \nonumber\\
\hat{S}_3 &=& \hat{n}_{\rm R} - \hat{n}_{\rm L} = -i \left( \ahd \av - \avd \ah \right), \nonumber
\end{eqnarray}
where $\hat{a}^{\dagger}_i$, $\hat{a}_i$ and $\hat{n}_i$ are the creation, annihilation and number 
operators for the corresponding modes. 
The average of these parameters, ${\bf S} = (\langle \hat{S}_1 \rangle, \langle \hat{S}_2 \rangle, \langle \hat{S}_3 \rangle)$, is ${\bf S} = (0, N, 0)$ for uncorrelated photons and ${\bf S} = (0, (N+1)/2, 0)$ for the Yurke state, indicating that these vectors align with the $S_2$ axis of the Poincare sphere.

The squeezing property of the Yurke state can be described using $\hat{S}_1$ and $\hat{S}_3$.  
For $N$ uncorrelated photons, the noise for $S_1$ and $S_3$ is equivalent, $\Delta S_1 = \Delta S_3 = \sqrt{N}$ (figure 1b). 
On the other hand for the Yurke state, the noise of $S_1$ is suppressed as $\Delta S_1=1$ at the expense of increased noise $S_3$, $\Delta S_3 = \sqrt{(N^2 + 2N -1)/2}$ (figure 1a). 
A large number of parameters have been devised to quantify spin squeezing for various applications \cite{Ma2011}. To characterise this squeezing, we use the squeezing parameter, $\xi_S$, which is defined to be the ratio between the minimum uncertainty for directions orthogonal to {\bf S} \cite{Kitagawa1993}, where $\xi_S <1$ indicates reduced quantum noise below the shot-noise. For uncorrelated photons, $\xi_S = 1$. While for the Yurke state, $\xi_S$ is minimized along the $S_1$ direction with $\xi_S=1/\sqrt{N}$ which indicates strong squeezing for this choice of squeezing parameter.

\begin{figure}
\begin{center}
\includegraphics*[width=9cm]{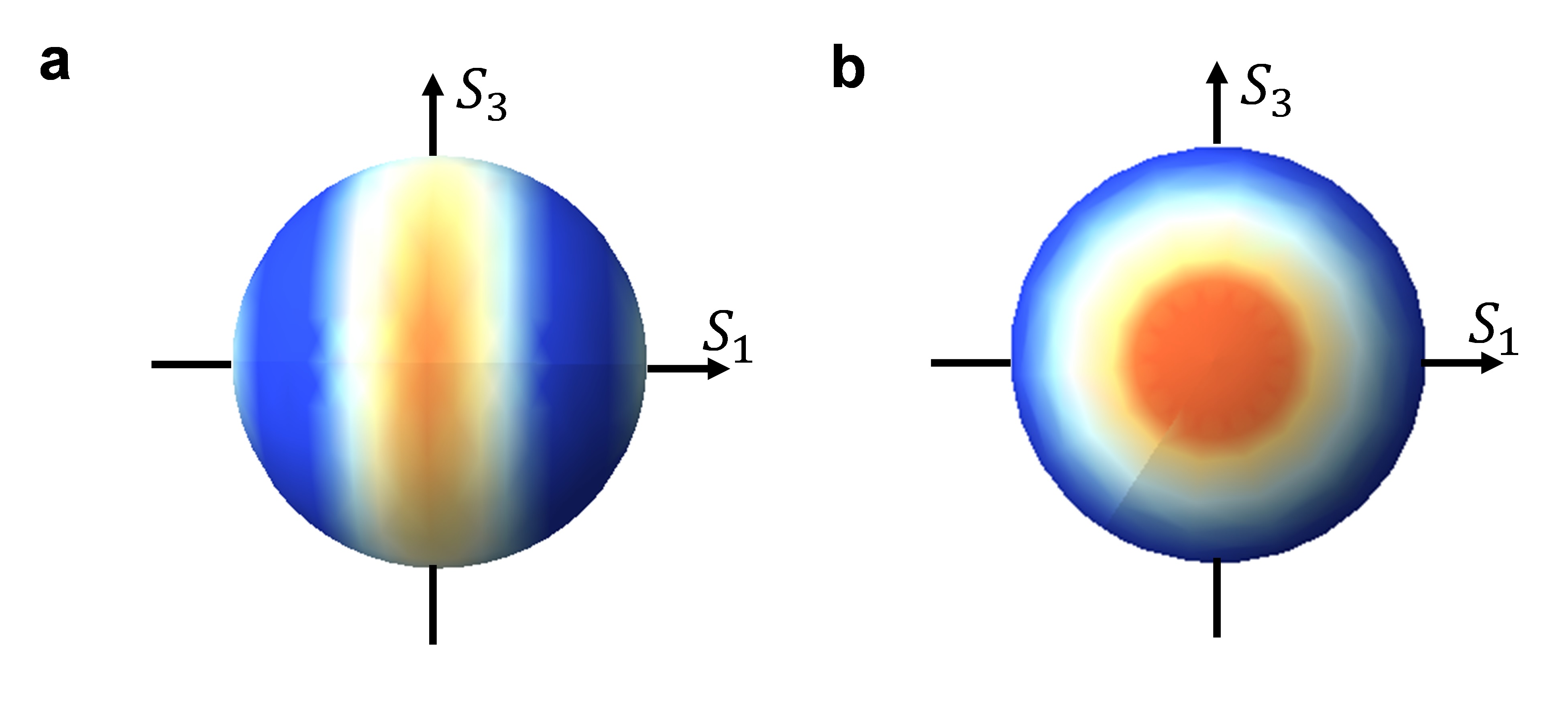}
\caption{\label{fig1}
{\bf Wigner distributions:}
The Wigner distributions are defined by average values for Stokes parameters. (a) corresponds to the five-photon Yurke state, (b) corresponds to five uncorrelated photons.   
}
\end{center}
\end{figure}

The squeezing property of the Yurke state can be used for improving the phase sensitivity of an interferometer.
The effect of a phase rotation by $\phi$ can be described by the unitary 
operator $\hat{U}(\phi) = \mathrm{exp}\left( -i \hat{S}_3 \phi/2 \right)$. 
Specifically, $\hat{S}_1$ after the phase rotation is expressed as 
$\hat{S}_1 (\phi) = \hat{U}^{\dagger} (\phi) \hat{S}_1 \hat{U} (\phi) = \cos \left( \phi \right) \hat{S}_1 - \sin \left( \phi \right) \hat{S}_2$.
Because $\langle \hat{S}_1 \rangle =0$ for both $N$ uncorrelated photons and the Yurke state, 
the average of $\hat{S}_1(\phi)$ is expressed by
\begin{equation}
\label{phase_s1}
\langle \hat{S}_1 \rangle (\phi) = -\langle \hat{S}_2 \rangle \sin \left( \phi \right).
\end{equation}
For estimates of $\hat{S}_1$, phase error is 
given by the ratio of $\Delta S_1$ and the phase derivative of $\langle \hat{S}_1 \rangle$.
Specifically, the phase error at $\phi=0$ is given by
\begin{equation}
\label{phase_sq}
\delta \phi_{sq} = \left. \Delta S_1 /\left| \partial \langle \hat{S}_1 \rangle /\partial \phi \right| \right|_{\phi=0} = \frac{\Delta S_1}{\langle \hat{S}_2 \rangle}. 
\end{equation}
Because $\Delta S_1 =1$ and $\langle \hat{S}_2 \rangle = (N+1)/2$ for the Yurke state, the phase error by squeezing of Yurke state is $\delta \phi_{sq}=2/(N+1)$.
To characterise the improvement of the phase sensitivity due to squeezing, we use another squeezing parameter $\xi_R$ which was introduced in ref \cite{Wineland1992,Wineland1994}, which is the ratio of the phase error for a general state and phase error due to shot noise $\delta \phi_{SNL}=1/\sqrt{N}$, with $\xi_R = \delta \phi_{sq}/\delta \phi_{SNL}$. For $\xi_R <1 $, the phase error is smaller than the shot-noise limit attained by uncorrelated photons $\xi_R =1$. For the Yurke state, $\xi_R= 2\sqrt{N}/(N+1) \approx 2/\sqrt{N} < 1$ for the high-$N$ limit.

Although squeezing of the Stokes parameters can improve phase sensitivity beyond the shot-noise limit,
 additional phase sensitivity can be achieved by phase estimation which accounts for the full distribution of measurement outcomes.
Statistical information about $\phi$ can be extracted from
the frequencies of every measurement outcome occurring in an experiment and
quantified using Fisher information $F(\phi)$. 
In a two-mode $N$-photon problem, Fisher information 
is calculated from the $N+1$ probability distributions, $p_{m}(\phi)$, where $m$ photons are detected with 
H polarization and $N-m$ photons with V polarization
\begin{equation}
\label{fisher}
F (\phi) = \sum_{m=0}^{N}{p_{m}(\phi) \left(\frac{\partial}{\partial \phi} \mathrm{ln}p_{m} (\phi) \right)^2}.
\end{equation}
More specifically, the Cram\'{e}r-Rao
bound states that any unbiased statistical estimator
of $\phi$ has mean-square error which is lower bounded
by $1/F(\phi)$, and this bound can be saturated using a suitable statistical estimator \cite{Casella2002}.
The minimum phase error for the Yurke state is then given by $\delta \phi_{opt} = 1/\sqrt{F}= 1/\sqrt{(N^2 + 2N -1)/2}$, where we used
$F = \Delta^2 S_3$ \cite{Hofmann2009}.
The improvement factor compared to the shot-noise limit is $\delta \phi_{opt.}/\delta \phi_{SNL} \approx \sqrt{2}/\sqrt{N}$ for the high-$N$ limit.
Note that the phase sensitivity obtained from the Fisher information is greater than the sensitivity obtained from equation (\ref{phase_sq}) 
by a factor of approximately $\sqrt{2}$, indicating that maximum sensitivity is achieved not only due to squeezing but also 
other quantum effects captured by the full set of measurement outcomes. Note that a theoretical analysis with similar motivation is given in \cite{Luca2009}.

%
\section{Experimental Setup.} 

\begin{figure}[b]
\begin{center}
\includegraphics*[width=9cm]{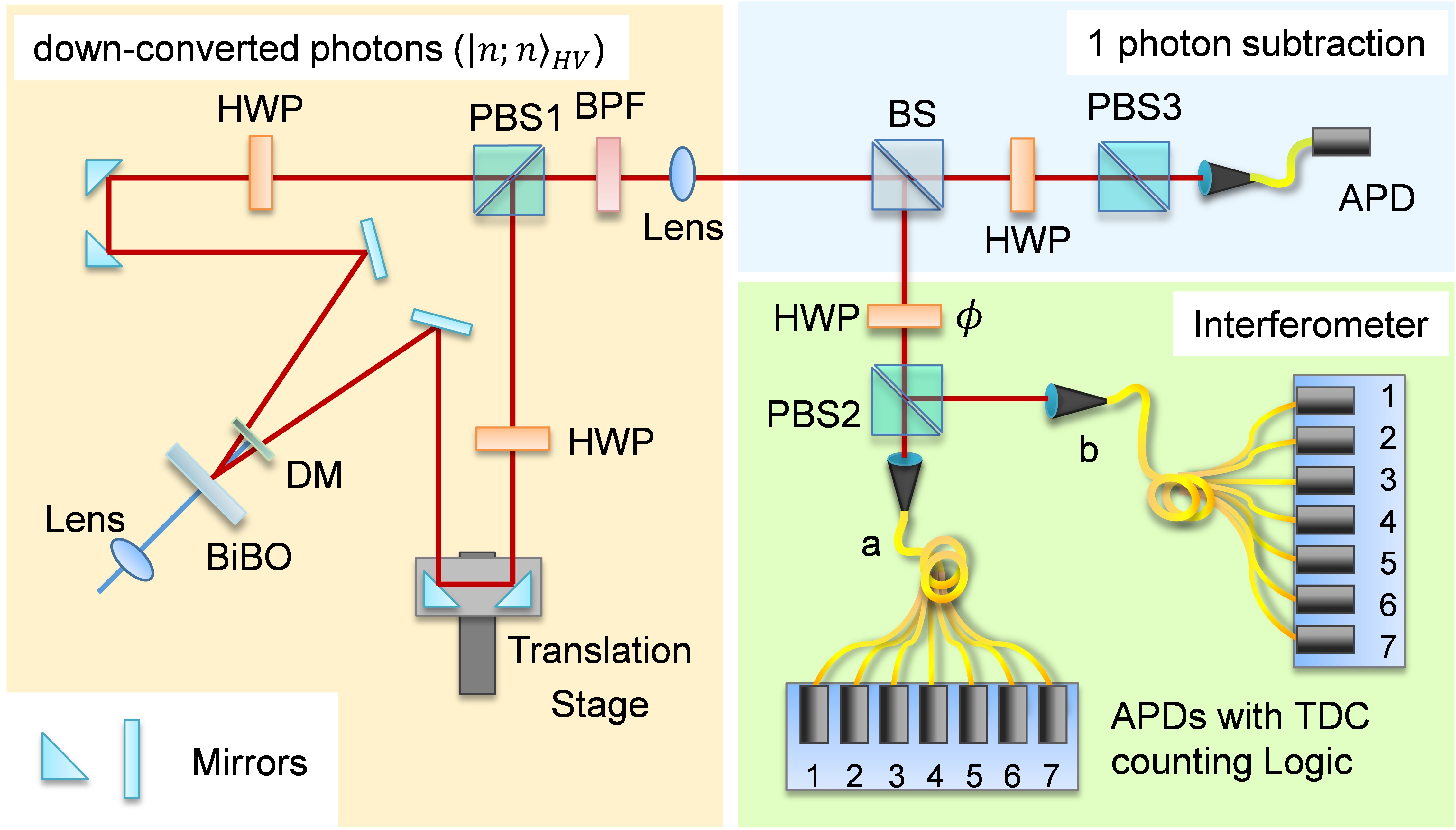}
\caption{\label{fig2}
{\bf Experimental set-up.}  Ultraviolet laser pulses, with a central wavelength of 404nm from a frequency-doubled mode-locked
Ti:sapphire laser (wavelength: 808nm, pulse width: 100 fs, repetition rate 80 MHz), pump a 2mm thick biaxial
Type-I bismuth borate (BiBO) crystal.  A dichroic mirror transmits the down-converted light and reflects the pump beam.
The signal and idler photons are rotated to 3 degrees with respect to the pump beam. The arrival timing between the signal and idler 
photons at PBS1 is adjusted by using a translation stage. After PBS1, a 3nm bandwidth band-pass filter is used to remove the fluorescent light.
}
\end{center}
\end{figure}

\begin{figure*}[t]
\begin{center}
\includegraphics*[width=18cm]{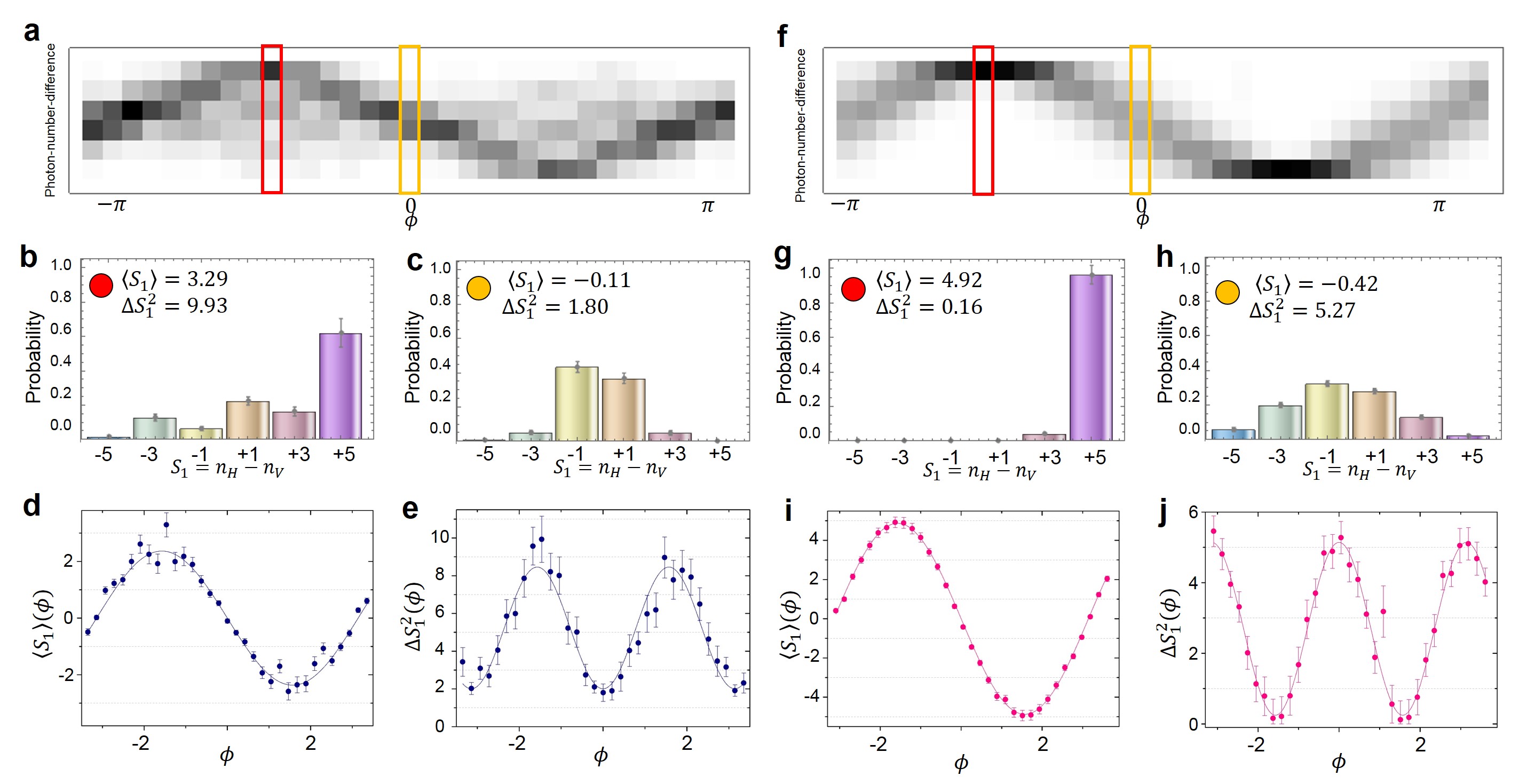}
\caption{\label{fig3}
{\bf Experimental results for the five-photon Yurke state (a-e) and the five-photon classical state (f-j) compared:} (a,f) are output photon-number distributions depending on phase shift $\phi$, (b,c) are output photon-number distributions for the Yurke state with bias phases $\phi=-1.46$ and $\phi=0.00$, (g,h) are output photon-number distributions for the classical state with bias phases $\phi=-1.63$ and $\phi=0.05$ respectively. (d,e) show the dependence of the average, $\langle \hat{S}_1 \rangle (\phi)$ and the noise, $\Delta S_1^2(\phi)$, on the bias phase at the output for the Yurke state, respectively. (i,j) show dependence of the average, $\langle \hat{S}_1 \rangle (\phi)$ and the noise, $\Delta S_1^2(\phi)$, on the bias phase at the output for the classical state, respectively.  
}
\end{center}
\end{figure*}
In order to demonstrate experimentally the phase sensitivity obtained from squeezing and maximum phase sensitivity
obtained from Fisher information, 
we have implemented phase measurement using a five-photon Yurke state.
Figure 2 shows the experimental setup for generating the Yurke state. 
Down-converted photon pairs are generated from biaxial Type-I bismuth borate (BiBO) crystal in a non-collinear configuration.
A half wave plate (HWP) is placed on each path so that one path is horizontally polarized and the other is vertically polarized. Each beam is then combined into a single spatial mode at the polarization beam splitter (PBS1).
The state after the PBS1 is a superposition of photon number states with equal photon number in the horizontal and vertical polarization, 
\begin{equation}
|\Psi_{\rm{PDC}} \rangle = \frac{1}{\cosh r} \sum_{N=0}^{\infty}{(\tanh r)^{N/2} |N/2, N/2 \rangle_{\rm{HV}}},
\end{equation}
where the sum is taken over even values of $N$.
If we postselect $N$ photons from this state, the state is equivalent to the Holland-Burnett state $|N/2, N/2 \rangle_{\rm{HV}}$\cite{Xiang2013}.

To generate the Yurke state\cite{Hofmann2006}, one photon is subtracted from the down-converted photon source, by detection of a single-photon in the D/A basis. After the one-photon subtraction, the conditional output is the five-photon Yurke state. In the setup, we put a beam splitter after PBS1 so that each of the $N$ photons in the beam is transmitted with probability 10\%. The transmitted one-photon state was measured in the D/A basis using a HWP set at $22.5$ degrees and PBS3. After the one-photon detection, the reflected $N-1$ photons are analyzed by the polarization interferometer.

To demonstrate the sub-shot noise phase measurement, we measured all possible coincidence outcomes, of which there are six, at the output as $\phi$ is varied. We used a pseudo-number-resolving multiplexed detection system using 1 $\times$ 7 fibre beam splitters, 14 avalanche photodiodes (APDs) and a multi-channel photon correlator (DPC-230, Becker $\&$ Hickl GmbH) \cite{Jonathan2013}. The phase shift was measured by using a HWP and PBS3 which were placed on the reflected path of the beam splitter.

%
\section{Results}
Figure 3 shows experimentally-obtained Yurke state interference (figure 3a-3e) and shot-noise limited interference (figure 3f-3j). 
Figure 3a and 3f show the effect on $S_1$ for the Yurke state 
and uncorrelated photons as $\phi$ is varied. As expected from equation (\ref{phase_s1}), the averages of the probability distributions follow the sine-pattern. The output noise in figure 3a is clearly reduced at around $\phi=0$ where the squeezing sensitivity is maximum, whereas the noise in figure 3f is limited by shot noise.
Figures 3b, 3c, 3g and 3h show the probability distributions for specific bias phases where the averages values of $S_1$ are nearly maximum (figures 3b and 3g) and are nearly zero (figures 3c and 3h) respectively. 
One can be seen from figure 3b that the effect of two photon coherence appears 
 as the photon-number oscillation for Yurke state \cite{Mehmet2009} (the peaks are observed at $S_1 = -3, +1$ and $+5$), and results in the 
reduced quantum noise at the phase where the average is nearly zero (figure 3c). 
On the other hand, for the classical uncorrelated case, the probability distribution of a classical state does not show the oscillation (figure 3g),
resulting in shot-noise at the output at the phase where the average is nearly zero (figure 3h).

\begin{figure}[t]
\begin{center}
\includegraphics*[width=8cm]{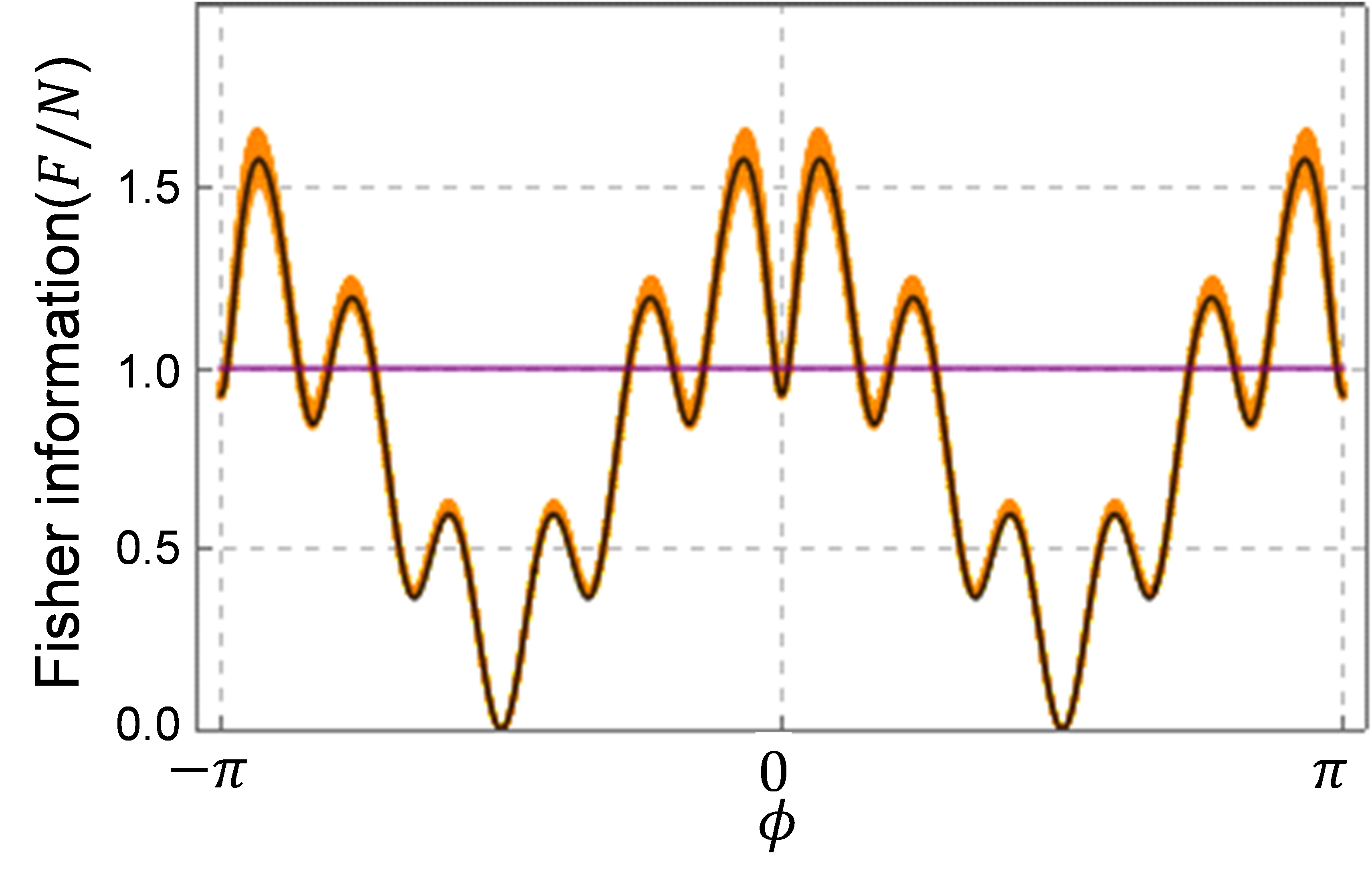}
\caption{\label{fig4}
{\bf Fisher information extracted from interference fringes:} The Fisher information for the Yurke state calculated using equation (3), from probability distributions
which are fitted to the experimental data.  The probability distributions are calculated using a model which
incorporates mode mismatch and noise, and the fitting uses rescaled detector counts  
 (see Methods section).  The purple line corresponds to the precision achievable at
the shot-noise limit.  The orange shading shows 200 iterations
of a Monte-Carlo simulation, for which the Fisher information is computed with Poissonian noise
added to raw detector counts.
}
\end{center}
\end{figure}

For more detailed analysis, figures 3d, 3e, 3i and 3j show the averages and noises for $S_1$ for the Yurke state and uncorrelated photons, respectively.
From the fitted curves, the phase derivative of the average output at $\phi=0$, 
$\left. \left| \partial \langle \hat{S}_1 \rangle/\partial \phi  \right| \right|_{\phi=0} = 2.37$ and the noise at $\phi=0$ is $\Delta S_1^2 (0) = 2.01$ for the
Yurke state. Thus, the phase sensitivity by squeezing is $\delta \phi_{sq} = 0.60$.
Similarly, $\left. \left| \partial \langle \hat{S}_1 \rangle/\partial \phi  \right| \right|_{\phi=0} = 4.87$ and $\Delta S_1^2 (0) = 5.14$ 
and $\delta \phi_{sq} = 0.47$.
The phase sensitivity obtained using the Yurke state here did not exceed the theoretical shot-noise limit of $\delta \phi_{SNL} = 1/\sqrt{5} \approx 0.45 $ even though the output noise of the Yurke state is smaller than the shot noise $\xi_S = 0.63$.

To extract the maximum phase sensitivity, we calculated the phase sensitivity by using Fisher information obtained from equation (3) for the Yurke state. Figure 4 shows the bias phase dependence of Fisher information.   
The maximum of Fisher information is $F=7.89$ at bias phase of $\phi = 0.21$ which is slightly different from the phase where the squeezing is maximum. Thus the obtained state can actually achieve sensitivity that is a factor of $1.58$ smaller than the shot-noise limit. 
Note that maximum Fisher information is obtained at a slightly-different bias phase where the squeezing is maximum. 
We can conclude that the improvement the phase sensitivity that we have demonstrated using the Yurke state is not only due to squeezing but also other 
quantum effects obtained by observing individual outcomes \cite{Strobel2015}.
In particular, Fisher information can extract the full information for changes in the phase parameter
 from the interference fringes at the output.

%
\section{Conclusions}
In conclusion, we have demonstrated, using our set-up, suppression of quantum noise by a factor of 2.56 with the effects of the squeezing being clearly shown by the measured interference fringes.
Spin squeezing is often characterised using parameters $\xi_S$ and $\xi_R$, where values $<$1 correspond to supra-classical performance.  Our measurements show clear spin-squeezing using the parameter $\xi_s= 0.63$, while our measurements of $\xi_R$, which is traditionally used to quantify sub-shot noise phase-noise error in spin-squeezing experiments, is $>$1.  None-the-less, 
the extracted Fisher information was 1.58 times better than shot-noise-limit demonstrating that quantum enhanced precision is possible even with $\xi_R \geq1$.
As an alternative to the multiplexed pseudo-number-counting detectors we used, recently-developed high-efficiency number-resolving detectors \cite{Takeuchi1999, Humphreys2015} could be used to improve detection efficiency and therefore reduce measurement time.
Our experimental demonstration is important not only for optical sub-shot-noise 
measurement but also other applications demonstrating sub-shot-noise spin-squeezed states \cite{Amico2008, Horodecki2009, Guehne2009}.

%
\section*{APPENDIX}
\subsection{Reconstruction of interference fringes for the Yurke state}
Photon-number counts at our multiplexed detectors are analysed as follows. Single photons are detected at each APD with probabilities of $\sigma_{a_i} (i=1, 2, ... 7)$ in mode $a$ and $\sigma_{b_j} (j=1, 2, ... 7)$ in mode $b$, which account for propagation loss and detector efficiency. 
In our analysis, we assume that five-fold coincidence detections arise only due to the generation of three photon pairs at the source (and neglect higher-order contributions). We define efficiency parameters for coincidence events at our multiplexed detectors as follows, where we assume that $m$ clicks in path $a$ and $5-m$ clicks in path $b$ correspond to $m$ actual photons in $a$ and $5-m$ actual photons in path $b$:
\begin{eqnarray}
\Sigma_m &=& \sum_{x_{a_1} +...+ x_{a_7}=m} {m! \sigma_{a_1}^{x_{a_1}}  ... \sigma_{a_7}^{x_{a_7}}} \nonumber\\
 & & \times \sum_{y_{b_1} + ...+ y_{b_7}=N-m}{ (5-m)! \sigma_{b_1}^{y_{b_1}}  ... \sigma_{b_7}^{y_{b_7}} },
\end{eqnarray}
where the variables $\{ x_{a_i} \}$ and $\{ y_{b_j} \}$ take values $0$ or $1$. To experimentally characterize $\Sigma_m$, we measured all of $\sigma_{a_i}$ and $\sigma_{b_j}$ (see Table I). 
Five-fold coincidence counts, $D_m(\phi_i)$, were then rescaled to give corrected count rates,
\begin{equation}
D'_m(\phi_i) = \frac{D_m(\phi_i)}{\Sigma_m},
\end{equation}

\begin{table*}
\centering
\caption{detection probability.}
\label{detpro}
  \begin{tabular}{c c c c c c c c c c c c c c}
\hline
\hline
   $\sigma_{a_1}$ & $\sigma_{a_2}$ & $\sigma_{a_3}$ & $\sigma_{a_4}$ & $\sigma_{a_5}$ & $\sigma_{a_6}$ & $\sigma_{a_7}$ & $\sigma_{b_1}$ & $\sigma_{b_2}$ & $\sigma_{b_3}$ & $\sigma_{b_4}$ & $\sigma_{b_5}$ & $\sigma_{b_6}$ & $\sigma_{b_7}$ \\ 
\hline
   1.40 \% & 1.25 \% & 1.43 \% & 1.46 \% & 1.53 \% & 1.54 \% & 1.48 \% & 1.16 \% & 1.45 \% & 1.30 \% & 1.12 \% & 1.11 \% & 1.36 \% & 1.58 \% \\
\hline
  \end{tabular}
\end{table*}

\begin{figure*}[t]
\begin{center}
\includegraphics*[width=18cm]{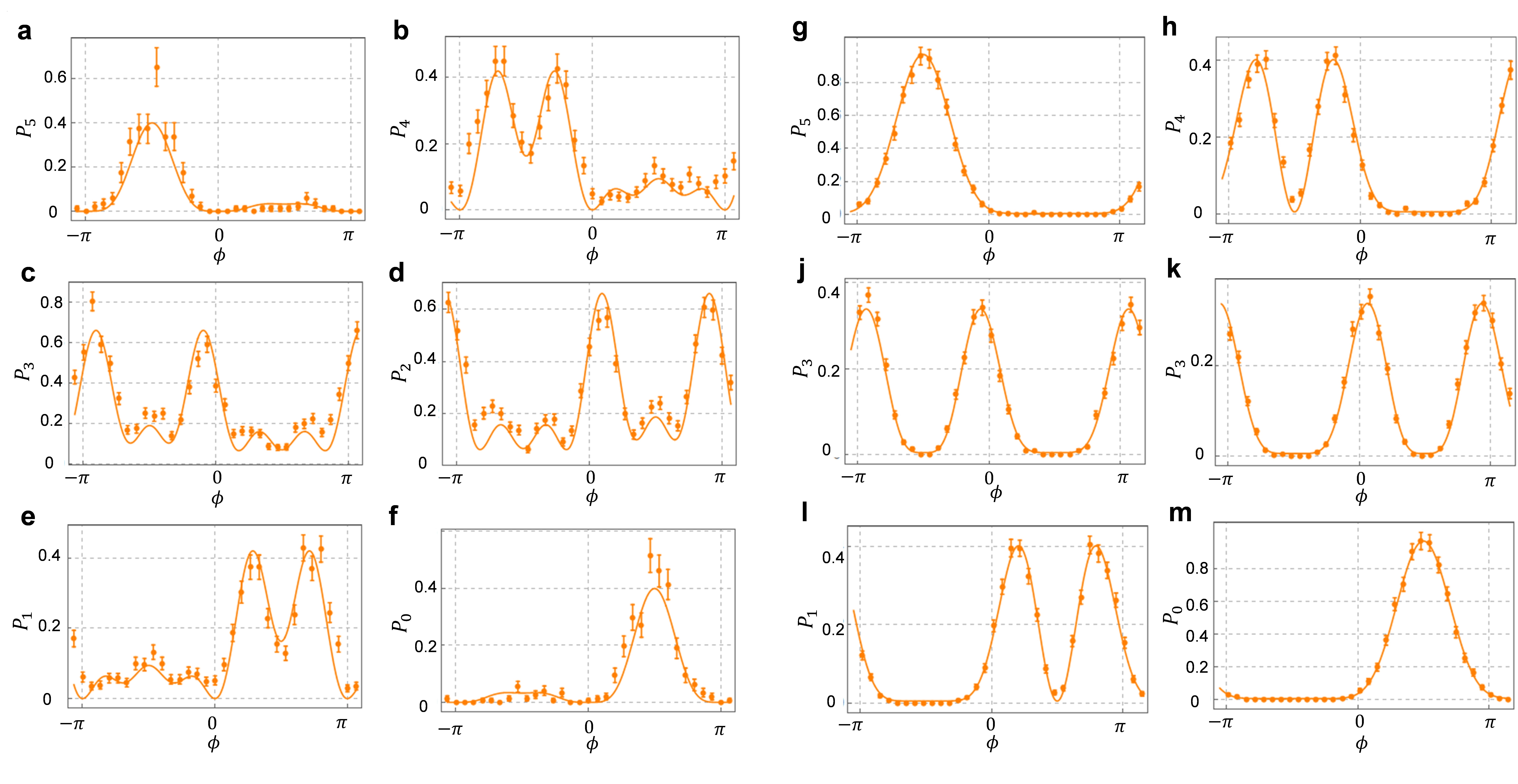}
\caption{\label{fig5}
{\bf Interference fringes for the five-photon Yurke state and the classical state.} 
(a-f) show the phase dependence of the output
probability distributions for the Yurke state. Data collection took eight hours per point. (g-l) show the phase dependence of the output probability distributions for the classical state. Data collection took one hour per point. $\phi$ was changed in intervals of $\pi/15$.
}
\end{center}
\end{figure*}

To implement the data fitting, we use theoretically-derived probability distributions $P_{m,I}(\phi)$ which model mode-mismatch, where $I$ is
mode overlap between horizontal and vertical polarizations (see below). Our fittings minimize
\begin{equation}
\sum_{m=0}^5{ \left( M \times P_{m,I}(\phi) - D'_m(\phi_i) \right)^2}
\end{equation}
using parameters $\phi$, $I$ and $M$, where $M$ is a scaling parameter.

\subsection{Derivation of probability distributions for the Yurke state including temporal mode mismatch.}
To derive $P_{m,I}(\phi)$, we start from theoretical model in \cite{Ra2013}. The quantum state generated before the BS in figure 2 is given by 
\begin{equation}
\label{psi}
|\psi \rangle = \frac{\left( \hat{a}^{\dagger}_{\rm {H}} \right)^{(N+1)/2}}{\sqrt{((N+1)/2)!}} \frac{\left( \hat{b}^{\dagger}_{\rm{V}} \right)^{(N+1)/2}}{\sqrt{((N+1)/2)!}} | 0 \rangle.
\end{equation}
$\hat{b}^{\dag}_V$ can be written as a superposition of one indistinguishable and one distinguishable component:
\begin{equation}
\label{indis}
\hat{b}_{\rm{V}}^{\dagger} = \sqrt{I} \hat{a}_{\rm{V}}^{\dagger} + \sqrt{1-I} \hat{a}_{\rm{V_{\perp}}}^{\dagger}, 
\end{equation}
where $I$ is the indistinguishability given by $\left| \langle 0 | \hat{a}_{\rm{H(V)}} \hat{b}_{\rm{H(V)}}^{\dagger} | 0 \rangle \right|^2 $, and  the symbol $\rm{V_{\perp}}$ denotes the orthogonal mode to $\rm{H}$ and $\rm{V}$. 
In the following, we assume that modes $\hat{a}_{\rm{H(V)}}$ and $\hat{a}_{\rm{H_{\perp}(V_{\perp}})}$ do not interact so that we can consider reduced density matrix, $\hat{\rho}_{|\psi \rangle}$ where offdiagonal terms can be neglected as follows,
\begin{widetext}
\begin{equation}
\label{rho}
\hat{\rho}_{|\psi \rangle} = \sum_{d=0} ^{(N+1)/2}{C_d |(N+1)/2 , (N+1)/2-d \rangle \langle (N+1)/2, (N+1)/2-d |_{\rm{HV}} \otimes |d \rangle \langle d |_{\rm{V_{\perp}}}},
\end{equation}
\end{widetext}
where $C_d$ is given by 
\begin{equation}
C_d = \binom{(N+1)/2}{d} I(\tau)^{(N+1)/2-d} (1-I(\tau))^d.
\end{equation}

Replacing the annihilation operators for indistinguishable and distinguishable modes as
\begin{eqnarray}
\hat{a}_{\rm{D}} &=& \frac{1}{\sqrt{2}} \left( \hat{a}_{\rm{H}} + \hat{a}_{\rm{V}} \right)\nonumber\\
\hat{a}_{\rm{D_{\perp}}} &=& \frac{1}{\sqrt{2}} \left( \hat{a}_{\rm{H_{\perp}}} + \hat{a}_{\rm{V_{\perp}}} \right),
\end{eqnarray}
where $\rm{D}$ denotes diagonal polarization, the state after the one-photon subtraction is given by a mixture of terms as follows,
\begin{equation}
\hat{\rho}_{I} = \frac{2}{N+1} \left( \hat{a}_{\rm{D}} \hat{\rho}_{|\psi \rangle} \hat{a}_{\rm{D}}^{\dagger} + \hat{a}_{\rm{D_{\perp}}} \hat{\rho}_{|\psi \rangle} \hat{a}_{\rm{D_{\perp}}}^{\dagger} \right). 
\end{equation}
The first term expresses the $(N+1)/2+1$ different distinguishability types and the second term expresses the $(N+1)/2$ different distinguishability types.

After the polarization interferometer, the state is transformed as $\hat{\rho}_{I}(\phi) = \hat{U}(\phi) \hat{\rho}_{I} \hat{U}^{\dagger}(\phi)$ where $\hat{U}(\phi)$ is unitary transformation due to a half-wave plate, which is expressed as 
\begin{eqnarray}
\hat{U}(\phi) \hat{a}_{\rm{H(H_{\perp})}}^{\dagger} \hat{U}(\phi) &=& \cos (\phi/2) \hat{a}_{\rm{H(H_{\perp})}}^{\dagger} + \sin (\phi/2) \hat{a}_{\rm{V(V_{\perp})}}^{\dagger} \nonumber\\ 
\hat{U}(\phi) \hat{a}_{\rm{V(V_{\perp})}}^{\dagger} \hat{U}^{\dagger}(\phi) &=& -\sin (\phi/2) \hat{a}_{\rm{H(H_{\perp})}}^{\dagger} + \cos (\phi/2) \hat{a}_{\rm{V(V_{\perp})}}^{\dagger} \nonumber\\
\end{eqnarray}
The probability that $N-m$ photons are detected in horizontally polarized mode and $m$ photons are detected in vertically polarized mode is then given by 
\begin{widetext}
\begin{equation}
\label{prob_mode}
P_{m,I}(\phi) = \sum_{d=0}^{n} \sum_{s=0}^{d} { \langle N-m-s , m-(d-s)|_{\rm{HV}} \otimes \langle s , d -\! s |_{\rm{H_{\perp}V_{\perp}}}~ \hat{\rho}_{I} (\phi) | N-m-s,m-(d-s) \rangle_{\rm{HV}} \otimes | s ,d-s \rangle_{\rm{H_{\perp}V_{\perp}}}}.
\end{equation}
\end{widetext}
which accounts for distinguishability with between 0 and n photons in the temporally-mismatched modes. As shown in figure 5, the probability functions given by equation (\ref{prob_mode}) fit in accordance to the experimentally obtained data.

\subsection{Calculation of Fisher information including phase insensitive noise.}
In spite of the accuracy of our theoretical model, some features in the interference fringes shown in figure 5 are not fully explained by the theory.  Care must be taken as estimates of $F$ are sensitive to perturbations in the fringes where there are extrema \cite{Jonathan2013}, as occurs for our experiment around $\phi=0$.  To ensure our estimates of $F$ are robust (and do not overestimate the true value), we add a phase-insensitive noise to the functions $P_{m,I}(\phi)$ as follows,
\begin{equation}
\label{prob_final}
P_{m,I,s}(\phi) = (1-s) \times P_{m,I}(\phi) + \frac{s}{6},
\end{equation}
where $m$ takes from $0$ to $5$.
Fisher information is then calculated using
\begin{equation}
F (\phi) = \sum_{m=0}^{5}{P_{m,I,s}(\phi) \left(\frac{\partial}{\partial \phi} \mathrm{ln}P_{m,I,s}(\phi) \right)^2}.
\end{equation}
$F$ in Fig. 4 is computed using these modified distributions, which have lower values compared to the unmodified distributions around $\phi=0$.


\section*{ACKNOWLEDGMENTS}
The authors would like to thank X. Q. Zhou and P. Shadbolt for construction of the experimental set up. 
The authors would also like to thank T. Stace, P. Birchall, and W. McCutcheon for useful discussion.
This work was supported by EPSRC, ERC, PICQUE, BBOI, US Army
Research Office (ARO) Grant No. W911NF-14-1-0133, U.S.
Air Force Office of Scientific Research (AFOSR) and the
Centre for Nanoscience and Quantum Information (NSQI).
J.L.OB. acknowledges a Royal Society Wolfson Merit Award
and a Royal Academy of Engineering Chair in Emerging
Technologies. J.C.F.M. and J.L.O’B acknowledge fellowship
support from the Engineering and Physical Sciences Research
Council (EPSRC, UK). J.S.C. acknowledges support from the
University of Bristol.

\end{document}